\documentstyle[12pt,aas2pp4]{article}
\def\insfig#1{#1}
\def\endinsfig{\end{document}}

\font\smallrm=cmr8

\def\FWHM{{\smallrm FWHM}}
\def\PSF{{\smallrm PSF}}
\def\RMS{{\smallrm RMS}}

\lefthead{Tonry, Burke, \& Schechter}
\righthead{The OTCCD}

\begin{document}

\title{The Orthogonal Transfer CCD\altaffilmark{1,2}}

\author{John L. Tonry}
\affil{Institute for Astronomy, University of Hawaii, Honolulu, HI 96822}
\affil{Physics Dept., Massachusetts Institute of Technology,
Cambridge, MA 02139}
\affil{Electronic mail: jt@avidya.ifa.hawaii.edu}
\authoremail{jt@avidya.ifa.hawaii.edu}

\author{Barry E. Burke}
\affil{Lincoln Laboratory, Massachusetts Institute of Technology,
Lexington, MA  02173}
\affil{Electronic mail: bburke@ll.mit.edu}
\authoremail{bburke@ll.mit.edu}

\author{Paul L. Schechter}
\affil{Physics Dept., Massachusetts Institute of Technology,
Cambridge, MA 02139}
\affil{Electronic mail: schech@achernar.mit.edu}
\authoremail{schech@achernar.mit.edu}

\altaffiltext{1}{Observations in part from the Michigan-Dartmouth-MIT 
(MDM) Observatory.}
\altaffiltext{2}{This work was sponsored by the Department of the Air
Force and the University of Hawaii Consortium.  Opinions,
interpretations, conclusions, and recommendations are those of the
author and not necessarily endorsed by the United States Air Force.}

\begin{abstract}

We have designed and built a new type of CCD that we call an
orthogonal transfer CCD
(OTCCD), which permits parallel clocking horizontally as well as
vertically.  The device has been used successfully to remove image
motion caused by atmospheric turbulence at rates up to 100~Hz, and 
promises to be a better, cheaper way to carry out image motion correction
for imaging than by using fast tip/tilt mirrors.  We report on the
device characteristics, and find that the 
large number of transfers needed to track image motion does not
significantly degrade the image either because of charge transfer
inefficiency or because of charge traps.  For example, after 100~sec
of tracking at 100~Hz approximately 3\% of the charge would diffuse
into a skirt around the point spread function.  Four nights of data at
the Michigan-Dartmouth-MIT (MDM) 2.4-m telescope also indicate that
the atmosphere is surprisingly benign, in terms of both the speed and
coherence angle of image motion.  Image motion compensation improved
image sharpness by about 0.5\arcsec\ in quadrature with no degradation
over a field of at least 3 arcminutes.
\end{abstract}

\section{Introduction}

Light from a distant object passing through the atmosphere suffers
phase distortions which cause an image of the object to dance and
blur.  These distortions are caused by turbulent density fluctuations
in the air which translate to fluctuations in the index of refraction.
The fluctuations are usually thought of as being stationary in time
but blown across a telescope aperture by the wind, causing the
wavefront of a point source to become a rapidly varying fractal
landscape.

The Kolmogorov (1941) theory of turbulence has proved to be quite
successful in describing how turbulence in the atmosphere
leads to the temperature, density, and index of refraction variations
which distort images.  A classic text is Tatarski (1961), Parenti
and Sasiela (1994) give a fairly recent and complete exposition, and
Beckers (1993) provides an overview of adaptive optics.
Briefly, turbulence of high enough Reynolds number has a self-similar
behavior and transfers energy from an ``outer scale'' characteristic
of the size where energy is fed into turbulent motions to an ``inner
scale'' which is small enough that motions on that scale lose their 
energy to viscosity in a
turnover time.  Scaling arguments then require that the structure
function $D_n$ of index of refraction variations over this range of scales
be given by
$$ D_n = \langle(n(\vec r_1)-n(\vec r_2))^2\rangle = C_n^2(h)\,r^{2/3},$$
where $C_n^2$ has units of m$^{-2/3}$ and depends on height $h$ above
the ground, but locally the structure function depends only on the
separation $r$ of the points if the turbulence is isotropic and
homogeneous.

The fact that the structure function increases rapidly with separation
has two consequences.  First, on a small enough scale (called $r_0$)
the \RMS\ distortions of the wavefront become less than a wavelength,
and $r_0$ can be thought of as the largest aperture which is
diffraction limited (albeit with substantial image motion) for given
seeing conditions.  The second is that without compensation there is
virtually no improvement in image quality as a function of telescope
aperture, since the contributions from the many subapertures of size
$r_0$ have no long-term phase coherence.

There are many ways to characterize image quality; one of the most
common and robust is the full width half maximum (\FWHM) of a point
source.  Typical seeing might have a \FWHM\ of 1\arcsec\ at
5000\AA, corresponding to an $r_0$ of about 10~cm.  When an
aperture is larger than $r_0$, the light from all the $r_0$ sized
subapertures interferes at the focus to produce a rapidly changing 
speckle pattern whose overall size corresponds to the diffraction
pattern of a single $r_0$, but which carries fringes and speckles
corresponding to the diffraction limit of the entire aperture.

Correcting the wavefront to recover the diffraction limited
performance of an aperture is known as adaptive optics, and it
requires phase corrections for each $r_0$ sized subaperture of the
telescope, on time scales which are roughly the time it takes the wind
to blow a distance of $r_0$, i.e. milliseconds.  The
difficulty is knowing the phase error of each subaperture, which
requires we collect perhaps 100 photons per subaperture per correction
time.  This is evidently independent of telescope aperture, and
steeply dependent on $r_0$.  A star of $m=7.5$ issues about 1 photon
per \AA\ per second per cm$^2$, hence we require a star (natural or 
artificial) of 
\begin{equation}
m \approx 8 + 7.5 \log({r_0\over10\hbox{cm}}) - 2.5 \log({v\over25\hbox{m/s}}),
\end{equation}
assuming a 3000\AA\ bandpass, a 50\% quantum efficiency, two samples
per transit time, and no detector noise.

A related problem is that of the ``isoplanatic angle,'' the angle over
which phase distortions are substantially the same.  In the context of
this simple model we might expect this to be $\theta \approx r_0 / h$,
where $h$ is the altitude where the phase distortions are taking
place.  This angle is small: for $r_0 = 10$~cm and 
$h = 2$~km, $\theta = 10$\arcsec.  If the distortions are all
occurring at a single altitude it is possible to conjugate them onto
the device making the phase corrections and thereby greatly increase
this isoplanatic angle, but this is seldom the case.

Finding a natural star which is bright enough and near enough to a
target of interest is such a demanding requirement that attention
has turned to the possibility of laser beacons, adaptive optics in the
infrared (the index of refraction becomes slightly smaller but mostly
the phase distortions are fewer wavelengths, hence $r_0$ becomes
dramatically larger, perhaps as large as a meter at 2$\mu$m), or
incomplete correction of phase distortions by correcting over a scale
much larger than $r_0$.

This last option is the natural one to choose when we can profit from
any improvement in image sharpness, when we require observations in
the optical, and when we have a finite budget and lifetime.  There are
a number of projects which fit this description; two which we have been
involved in are surface brightness fluctuations where the darker
skies and quality of optical detectors are important (see Jacoby et
al. 1992 or Tonry et al. 1997 for a review), and photometry
of close gravitational lenses to search for time delays, where the
fluctuations in QSO luminosity are greater in the optical (e.g.
Schechter et al. 1997).  

The usual approach in this incomplete correction of the wavefront is
to decompose the wavefront into modes, usually Zernicke polynomials
which are orthogonal polynomials on a circular disk, and correct for
some limited number of modes.  Effectively what we are doing is
deliberately using an $r_0$ in the above formula which is just as
large as we please, knowing that we will not be achieving diffraction
limited performance, but hoping that there will be a worthwhile
improvement in image quality.  Evidently, this can lead to
improvements in the requirements for brightness of guide
stars, the speed at which we have to track the motion, and the angle
over which the correction is effective.

For example, plugging $r_0 = 2.4$~m into the formulae above 
suggests that we should be able to work with $m = 18$ guide
stars, that the rate at which we will have to track the motion might be
20~Hz for a 25~m/s wind, and that we can expect a correlation angle
(usually known as the ``isokinetic angle'' when we are only talking
about full aperture image motion) of 4 arcminutes for phase
distortions at $h = 2$~km.

The lowest two modes we could consider are $x$ and $y$ gradients,
colloquially known as ``tip/tilt,'' and more precisely termed
``G-mode'' and ``Z-mode'' depending on whether we use an average of
the gradient values over the wavefront or fit a least-squares
gradient to the wavefront (the two lowest Zernicke terms).  These
modes correspond to image motion: the entire speckle pattern dances
around with some amplitude.  Since $r_0$ is defined as a subaperture
over which there is roughly one wavelength of wavefront distortion, 
we can expect that a
single subaperture will have a motion \FWHM\ of $\Delta\alpha_1 \approx
\lambda / r_0$, i.e. very similar to the \FWHM\ from diffraction.

If the structure function were not rising as a function of separation,
we would see a reduction in image motion inversely proportional to
telescope aperture $D$, because of the averaging of the positions of
the images from each subaperture.  However, the extra power at large
scales causes the image motion to decrease only as $D^{-1/6}$, 
i.e. it is always comparable to the diffraction size of a single $r_0$
subaperture.  Thus it is always important to remove the image motion
from the atmosphere as well as image motion from telescope shake.

There have been a number of instruments built to compensate for image
motion.  Some of the more successful have included ISIS (Thompson and
Ryerson 1984), the HR-Cam at the CFHT (McClure et al. 1989) and its
successor the MOS/SIS, the COME-ON AO system (which is capable of
higher order correction than just tip/tilt) at ESO's NTT (Rigaut et
al. 1991), and various tip-tilt secondaries (Steward 90-in, UH 2.2-m,
OCIW 2.5-m at Las Campanas).
These work by incorporating a mirror into the light path
which can be rapidly tilted to steer the light, either a small
mirror in a reimaged beam, or the secondary of the telescope itself.
A novel approach carried out by McCarthy et al. (1997) is to
mount the detector itself on piezoelectric actuators and shake the
focal plane.  The challenge with all of these devices is to make the
moving element light and stiff enough to push resonances to high
enough frequencies that it is possible to maintain a stable control
loop at 20--50~Hz.  This is not an insignificant engineering
achievement and it is generally quite expensive to do, particularly
the most ambitious projects where the telescope secondary is articulated.

The gains from using these devices can be significant.  Not only can
image motion from the atmosphere be frozen, but also any telescope
shake can be removed.  Nevertheless, these devices are often difficult
to use, offer small gains, and cause enough degradation in images (loss of
light, non uniform \PSF, setup overhead) that most images taken
today do not benefit from image motion compensation, other than using
a standard autoguider sampling at rates less than 0.5--1~Hz.

We have embarked on a project to achieve image motion compensation in
a somewhat different way, by moving the electrons within a CCD to
follow the optical image falling on the device.   This is a new type
of CCD, which we call an orthogonal transfer CCD (OTCCD), designed and
built at MIT Lincoln Laboratory, which substitutes a fourth gate for the
usual channel stop, hence permitting side-to-side parallel clocking of
charge as well as up and down.  (Ken Freeman has coined the acronym
MIAOW for minimum inertia adaptive optics widget.)  Our first
device was only $64\times64$, but demonstrated that the concept works
and that it is possible to shift a developing image within a CCD to any
position (Burke et al. 1994).  We have since built a $512\times512$
device with equal size frame-store region, and we have collected
engineering and scientific results at the telescope with it.  The
advantages of this device are that it can shift charge {\it much}
faster than an optical image is likely to move (within $\approx
100\mu$sec), it suffers from none of the problems incurred with
moving optics, and it is an extremely cheap option to build into a
CCD.

The next section will describe the device in more detail, along with
how it is actually used at the telescope.  The third section will
discuss the main drawback to this device: pockets and traps create
holes in images when the charge has been shifted back and forth many
times.  The fourth section will present results from four nights of
use in July 1996 at MDM Observatory.  We report on the significant
improvements in image quality using this device.  We will then conclude
and speculate on directions for its future development.

\section{Structure and Use of the OTCCD}

Figure 1 
\insfig{
\begin{figure}
\epsscale{1.0}
\plotone{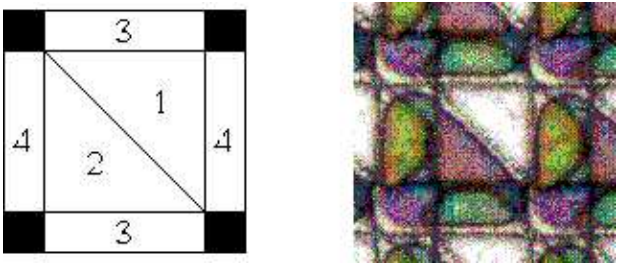}
\caption[fig1.eps]{
On the left is a schematic of how the four gates of an OTCCD are
organized.  On the right is a photomicrograph of an OTCCD pixel.  In
this device gate \#1 is aluminum and therefore appears bright.
\label{fig1}}
\end{figure}
}
shows a schematic diagram of the gates overlaying a pixel in
the OTCCD, and a photomicrograph of an actual device.  In a normal CCD
the fourth phase would be a channel stop implant, whereas in the OTCCD
the channel stops are small implants at the corner of each pixel.  If
this fourth phase is set low, however, it acts like a conventional
channel stop and we can clock the charge in the device down with the
3-phase sequence 1--2--3--1 or up with the sequence 1--3--2--1.
However, we have also altered the geometry of the two central phases,
1 and 2, into triangles which are rotation symmetric, and the third
phase is symmetric with phase 4, so that we can also clock charge left
and right.  This is achieved by setting phase 3 low and using the
sequences 1--4--2--1 for right movement or 1--2--4--1 for left
movement.  For normal integration we set both phase 3 and phase 4
negative and collect under phases 1 and 2.

The first realization of this geometry was a $64\times64$ test device
which worked as expected (Burke et al. 1994), but was
too small for effective tests of charge transfer inefficiency (CTI) or
rate of incidence of pockets and traps.  Fortunately, the chords 
of a mask set for another CCD development program became available at
this time, and we 
seized the opportunity to build a larger OTCCD.  (The other program
only used the square inscribed within the circular wafers.)

In order to make a device which could track image motion without any
external inputs, we decided to build a chip with a $512\times512$
array of OT gates and a $512\times512$ array of conventional 3-phase
gates.  This lower portion of the chip could be used as a frame-store
area, but we chose to leave it uncovered so that it could be used to
track a guide star.  Both arrays dump into a split serial register
with amplifiers at each end.  We wanted this CCD to be as simple as
possible to use, so we kept the number of clocks to 12 so that the
OTCCD could be run with a single SDSU/Leach analog board (4 OT gates,
3 three-phase gates, 3 serial gates, a summing well, and reset).  The
pixels are 15~$\mu$m which translates to 0.17\arcsec\ at the 
Michigan-Dartmouth-MIT (MDM) 2.4-m
telescope, and the entire chip is therefore 1.5\arcmin$\times$3\arcmin.

Because of financial constraints and because we did not want to
introduce any processing which might jeopardize the standard
three-phase devices on the same wafer,
we chose to use aluminum for the topmost gate (number 1).  As
we discovered, this may have increased the incidence of pockets.

Figure 2 
\insfig{
\begin{figure}
\epsscale{0.8}
\plotone{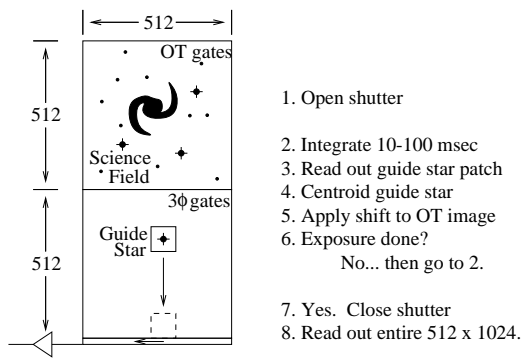}
\caption[fig2.eps]{
The OTCCD is used to track image motion by placing the science field
on the portion of the CCD with OT gates and a guide star on the
frame-store portion of the CCD.  A subarray containing the star image
is rapidly read out and used to correct the position of the charge
within the OT array to follow the motion of the science field.
\label{fig2}}
\end{figure}
}
illustrates how the OTCCD is operated at the telescope.  The
telescope is first positioned to place the field of interest on the OT
region of the chip.  Next, a suitably bright guide star is sought,
possibly by rotating the instrument.  For the observations in July we
were using a front-side illuminated device (where aluminum covers
about 60\% of the chip), so the quantum efficiency was low.  A star of 
$m_I \approx 23.5$ gave us 1$e^-$/sec through a standard Kron-Cousins
$m_I$ band filter, and we could track at 10~Hz on
stars of about $I = 15.5$.  The probability of finding a star within the
1.5\arcmin\ field at this brightness is about 0.6 at the north
galactic pole, so for some fields we had to track more slowly than
10~Hz.  The back-side illuminated devices currently being fabricated
will improve this by about 1.5 magnitudes and we can
improve the signal to noise of our guide star detection by about a
factor of 2 (another 0.7 mag), which will bring the probability of
finding a guide star for 10~Hz within 1.5\arcmin\ to greater than 0.9.  

We use our technique of ``shutterless video'' (Metzger, Tonry, and
Luppino 1993)
to read out guide star information while the OTCCD is integrating on
the science field.  This method is applicable to any CCD, whether or
not it has a shadowed frame store area (the OTCCD does not), and it is
the usual means at MDM to focus the telescope and examine out of focus
donuts for miscollimation.  The shutterless video and the resulting image
tracking loop are illustrated in Figure 2.

This method has some shortcomings compared to a device with a real
frame store: there is a bit of vertical streaking from the star as the
charge is clocked past, the guide star subarray must be at least one
subarray size above the serial register, it is possible to have ghosts
if there happens to be a bright object above the serial register where
the guide star image is being read out, and there is some of loss of
duty cycle because the time while the image is read out ($\approx
4$~msec for a $16\times16$ patch) cannot be used for integration.
On the other hand, it is possible to select an arbitrary guide star
region with arbitrary binning, and this can be used with any CCD.

The choice of where to move the charge collecting in the OT region as
a result of the guide star information is an interesting exercise in
control theory.  In this case we have quite accurate information of
guide star location (mostly good to a fraction of a pixel) extending into
the past, but of course the information is always one sample time
late.  The device we are controlling (the electrons in the OT region) 
has ideal properties: there is no inertia or backlash and we carry out
the OT shifting in about 80~$\mu$sec per pixel (and a typical shift is
only one pixel).  Some algorithms which we have considered are
(1) prediction = previous position, (2) prediction = linear
extrapolation of previous positions, and (3) linear predictive coding
or Kalman filtering of the previous positions.  (Numerical Recipes by
Press et al. 1992 has a short section on LPC and references to some of
the standard engineering textbooks.)
Algorithm \#1 is 
pretty poor for a sinusoidal motion sampled at the Nyquist frequency,
but algorithm \#2 is even worse.  We do not have much experience with
algorithm \#3 yet; in the lab it usually does spectacularly well,
but is computationally expensive.
All of the observations presented here are algorithm \#1
with a ``veto'' limit on how large a shift a sample can call for.

Ideally we would like to optimize the tradeoffs between signal to
noise and tracking speed, but whatever algorithm we select must be
quite fast (computing in less than a few msec on a Sparc 2).  We also
require that the algorithm be extremely robust, because we only read
out a small patch around the guide star and adjust the location of
the subarray for the next sample
according to where we think the guide star will be, so we cannot
afford to have a bad sample or cosmic ray throw us off by more than a
subarray size.  This is quite a well defined problem in mathematics and
control theory, and we would welcome suggestions on optimal
algorithms.

As the observation proceeds, each of the guide star images is piped to
a display which also shows the current centroid, a leaky average of
the guide star images, and a strip chart of the image \FWHM.  One of
the nice features about the OT tracking is that if the telescope focus
drifts the observer can refocus in real time without worrying about
causing the image to move.

When the exposure has finished the entire CCD reads out and the
science image in the OT region is saved to disk.  We also save a
complete record of time, guide star position, and image shifts for
each guide sample.  This is necessary for creating a flat field
because the science image has been dithered over many OTCCD pixels.
If requested, the software will also record a 3-dimensional FITS file
of the actual guide star images, and this movie can be replayed or
analyzed for information other than centroid.

We have also written software to read one of these records of image
shifts created during an exposure and to replay it at arbitrary
speed.  We thought this would be a better way to make a flat field:
an image's actual shifts could be reproduced while the OTCCD is
viewing a flat field and the pocket-pumping would occur in the same
way.  As we learned, however, this is both ineffective and
unnecessary.  

Presently we use the OT guide star information purely for OT tracking,
and we depend on the usual telescope autoguider to guide the
telescope.  This is purely a matter of expediency and we can probably
improve performance a bit (as well as making the observer's job
simpler) by using a suitably filtered version of the OT guide star
motion to move the telescope.

Since we had only five weeks from the receipt of the first OTCCD
until our first observing run (during which we had to build a focal
plane, write DSP code to run the OTCCD, characterize chips, write DSP
code to do the OT shifting, write code for the Sparcstation to do the
centroiding, exposure management, data recording, and image display), 
there are a number of improvements we would like to carry out.  First
and foremost, we are far from extracting optimal signal to noise from
our guide star images.  When we can control the telescope position
from the Sparcstation we should be able to center and bin
appropriately to do quadrature detection.  The image display can be
made much more useful for the observer: a full frame view mode and
some sort of guide star subarray graphic would be helpful.  

The most interesting improvement we would like to implement is some
sort of fast auto-focusing.  A quadrature detection produces four
numbers, call them A, B, C, and D (Figure 3).
\insfig{
\begin{figure}
\epsscale{0.4}
\plotone{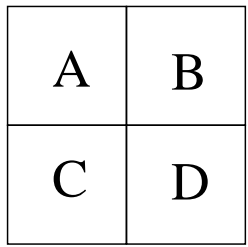}
\caption[fig3.eps]{The four pixels of a quadrant sensor.
\label{fig3}}
\end{figure}
}
We get flux information
from A+B+C+D, $x$ guide signal from (A+C)$-$(B+D), and $y$ guide
signal from (A+B)$-$(C+D).  If we deliberately introduce some
astigmatism at $45^\circ$ to the pixels, we can get a focus signal
from (A+D)$-$(B+C).  This is a well-known trick for CD players.  In
our case, by placing a cylindrical lens of the proper thickness and
power above the CCD we can place the two astigmatic foci symmetrically
about the nominal image plane so that the focus signal will be zero
when the telescope is in focus.  This focus signal could either be
filtered and sent to the usual telescope focus mechanism, or it could
be used to drive some fast focus adjustment, such as a tilting plate
or a piezo drive of the focal plane itself.

\section{CCD Performance and Pockets}

\subsection{Characterization in the Laboratory}

We used iron-55 xrays and adjusted output FET drain voltage, $V_{DD}$,
bias, and clock levels to optimize the CCD performance.  Figure 4
\insfig{
\begin{figure}
\epsscale{1.0}
\plotone{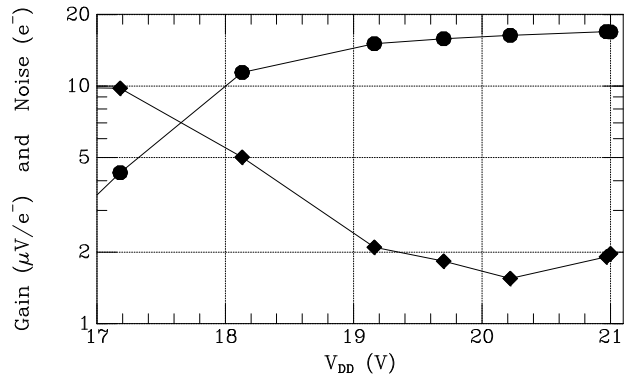}
\caption[fig4.eps]{
The read noise (declining curve) and responsivity (rising curve) of
the OTCCD output amplifier as a function of drain voltage.
\label{fig4}}
\end{figure}
}
shows the gain and noise level of the CCD as a function of $V_{DD}$ at a
frame rate of about 50~kpix/sec.  At $V_{DD}$ of about 20~V, the CCD
amplifier puts out about 16$\mu$V per electron and achieves 1.6$e^-$
noise.

A typical image taken with OT tracking will undergo 5000 shifts
(perhaps 20 per second over a 250~sec exposure).  We measure the
CTI of the OTCCD to be better than $3\times10^{-6}$, which means
that a charge packet might lose 1\% of its charge into a
surrounding halo during the course of an exposure.  This is not
a significant degradation of the point spread function.

The biggest concern about the viability of image motion compensation
within an OTCCD was that traps or pockets might cause serious
``pocket-pumping.''  For example, a bulk trap of only one electron
depth might store an electron during each shift and release it later,
potentially digging a 5000$e^-$ hole in the image and building up a
5000$e^-$ spike.  Since typical exposures have sky levels of only a
few hundred to a thousand electrons, this will seriously degrade the
image.  Early experiments by 
Stockman (1982) with RCA chips 
(and TI $800^2$ CCDs also suffered badly from traps)
indicated that this was a fatal problem for these devices unless very
few shifts were performed.  Modern
CCDs have far fewer traps and pockets, and have been used successfully
for image chopping (e.g. Cuillandre et al. 1994 or Sembach and Tonry
1996), but we did not know how serious the problem would be with the
OTCCD.

We experimented with deliberate pocket-pumping in the lab, either by
shifting back and forth many pixels (at about 80$\mu$sec per shift)
or by shifting back and forth by only two phases (to try to discover
the subpixel location of traps).  When the number of shifts was
small enough that there were still electrons left in the afflicted
pixel, we characterized the depth of the pocket as the depth of the
hole or equivalently the size of the adjacent spike, and express
the depth in units of electrons per clock cycle.
We found that the number of pockets
is quite variable from device to device, some having thousands of
pockets bigger than 1$e^-$, others with as few as a hundred.  Figure 5
\insfig{
\begin{figure}
\epsscale{1.0}
\plotone{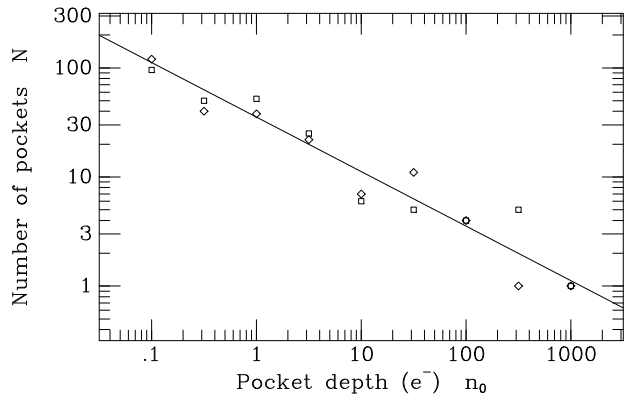}
\caption[fig5.eps]{
The number of pockets as a function of the pocket depth (determined by
pocket pumping at a rate of 80$\mu$sec per shift).  The squares
and diamonds refer to pockets which trap charge when it is clocked
up-down or left-right.
\label{fig5}}
\end{figure}
}
shows the distribution of pockets in one of the better devices.  The
number of pockets $N$ depends roughly on pocket depth $n_0$ as
$N \approx n_0^{-1/2}$.

The density of pockets in the OTCCD region of the best devices is
about three times higher than the 3-phase region on all the chips.
For temperatures cold
enough to have negligible dark current we saw little change in pocket
number and size with temperature.  
The pockets did become fewer (and smaller) as we 
clocked more slowly, but we needed a geometric increase in
parallel transfer time to obtain a linear improvement in pocket size.
The pockets are extremely sensitive to clock voltage: the
bigger the voltage difference between adjacent phases during clocking,
the smaller the pocket size.   We could set the parallel clocks as
negative as $-10$~V (although the surface potential pins around $-6$~V), and
we found that we could not set the parallels more positive than about
$+4$~V, without incurring some breakdown between the parallel and
serial clocks.  Apart from these general observations, we have not yet
carried out detailed, quantitative characterization of pocket size
as a function of all the various parameters.

A striking feature of these pockets is that they are virtually all
unidirectional, i.e. they block the passage of a bit of charge in one
direction, but do not impede charge in the other direction, a bit like
water flowing over a curbstone.
Generally speaking, the pockets trap charge when
clocking either up or to the right, although some (and many of the larger
ones) will trap charge when clocking down or left.

All of these considerations suggest that the pockets are
localized electrostatic potential minima.  The up-right (U-R) and
down-left (D-L)
unidirectionality of the trapping also points to something to do with
the $45^\circ$ interface between the 1 and 2 phases.  This will be
examined in more detail below.

\subsection{Performance at the Telescope}

The OTCCD performed well when we put it on the MDM 2.4-m
telescope.  What was surprising, however, was the fact that the
pockets were much less pronounced than we had feared.  Figure 6 
\insfig{
\begin{figure}
\epsscale{1.0}
\plotone{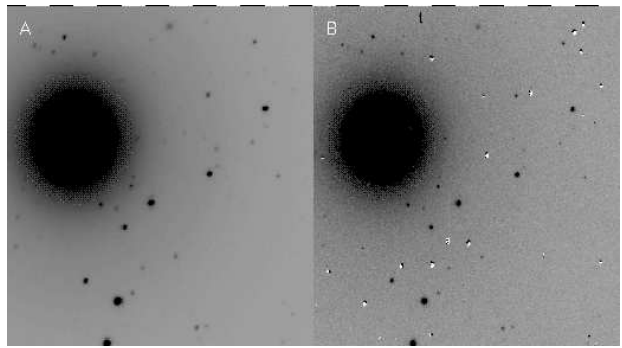}
\caption[fig6.eps]{
Two images of NGC~6703.  The left image (A) was taken with a Loral
CCD using normal guiding, and had an equivalent exposure of 5000~sec.
The right image (B) had 4650 shifts during a 500~sec
exposure.  The \PSF\ \FWHM\ is 0.82\arcsec\ on the left and
0.58\arcsec\ on the right.  The stretch is adjusted to make the
pockets stand out.
\label{fig6}}
\end{figure}
}
shows
an image of NGC~6703 at a contrast level to show 1\% deviations from the sky
level.  There about 20 pockets apparent, but the rest are nearly
invisible.  Figure 7 
\insfig{
\begin{figure}
\epsscale{1.0}
\plotone{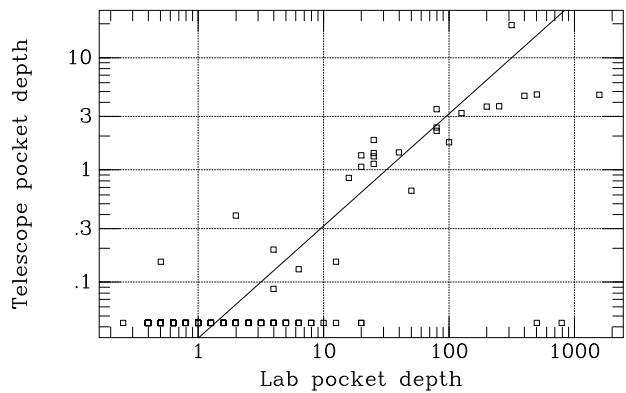}
\caption[fig7.eps]{
The size of pockets in images derived during actual observing is
plotted against the size of the same pockets derived from pocket
pumping experiments in the lab.  The line illustrates a factor of 30
reduction in size.
\label{fig7}}
\end{figure}
}
shows the size of the pocket (in $e^-$ per
transfer) measured in this image versus the pocket size measured in
the lab.  Each pocket which we had measured in the lab was about a
factor of 30 less prominent in astronomical images taken at the
telescope.  Since the image motion \RMS\ is about 2 pixels, the entire
pattern that the pocket traces is about 50 pixels.  This image
underwent 4650 shifts and the sky level is about 850$e^-$.  Therefore
a pocket which traps 3$e^-$ per shift in the lab (and appears to
trap only 0.1$e^-$ per shift here) would dig a
hole in this image of depth $4650\times0.1/50$ and just be visible
at a level of 1\% of sky, whereas without this diminution of pocket
size it would cause a deep hole of 30\% of sky.

Expecting that the pockets would create images that were littered with
pits, we wrote software to replay the pattern of shifts from an image
while the OTCCD was illuminated by a flatfield.  Any pocket deep
enough to dig down to the bias level could not be corrected, but we
thought that this would remove the smaller pockets.  It turned out,
however, that such flatfields do not to a good job of mimicking the
pocket patterns in the data.  Generally speaking, a flatfield is much
brighter than a sky image so the replay of the shift pattern happens
much more rapidly than the data exposure, and the eventual image is
also much brighter than the data.  Our realization of how sensitive
the pockets are to dwell time enables us now to understand why this
procedure is not effective in removing pockets.
It is conceivable that we could use these flatfield data and a
physical model to correct for the effects of the pockets, but they
comprise such a small fraction of the area (about a percent), that it
is probably more effective to simply to move the telescope between
exposures and delete the data affected by pockets.

\subsection{Diffusive Draining of Pockets}

We believe that our pockets are electrostatic potential minima, 
as discussed by Chatterjee et al. (1979).
The thermal diffusion of charge out of a pocket is described by
$J = -D \nabla n,$
where $J$ is the flux of electrons, $D$ is a diffusion constant which
depends on temperature, and
$\nabla n$ is the gradient of the number density.  If we approximate the 
pocket as a rectangular volume of width $L$ and cross section $A$, the
barrier width as $\Delta L$, the barrier height as $\Delta \phi_0$
when the pocket is empty and $\Delta\phi$ when the pocket is partially
filled with charge,
and the number of electrons in the pocket as $N$ (Figure 8),
\insfig{
\begin{figure}
\epsscale{0.5}
\plotone{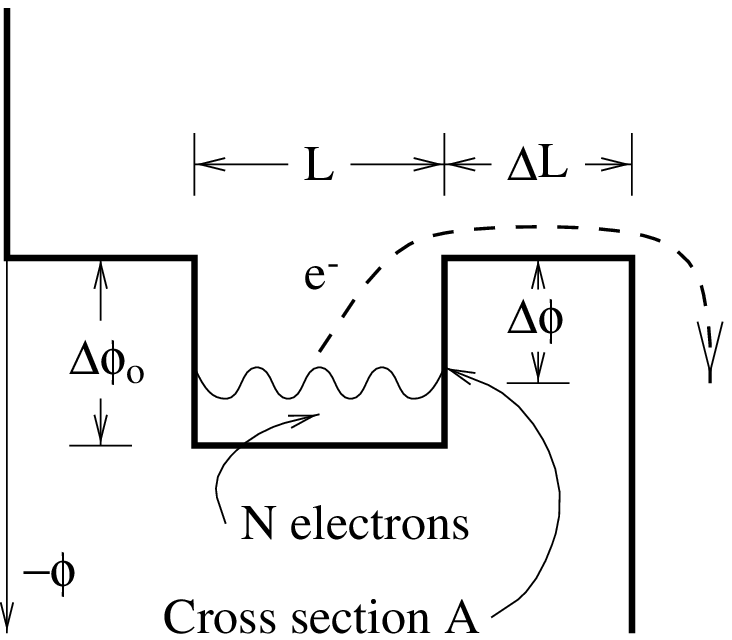}
\caption[fig8.eps]{
A pocket of volume $L * A$ has a potential barrier of
height $\Delta\phi_0$ when the pocket is empty and width
$\Delta L$.  Electons must overcome the barrier $\Delta \phi$ in order
to escape from the pocket.
\label{fig8}}
\end{figure}
}
we have $dN/dt \approx J A$, $\nabla n \approx N / (AL) / \Delta L$,
and $D \approx D_0 \exp(-e\Delta\phi / kT)$.  Hence
\begin{equation}
{dN(t)\over dt} = -{D_0 \over L \Delta L} N(t) \exp(-e\Delta\phi(t)/kT).
\end{equation}
We can also relate the barrier potential to the number of electrons in
the pocket via a capacitance: $\Delta\phi = e(N_0 - N) / C$, where
$N_0$ is the pocket capacity.  Thus, 
\begin{equation}
{dN(t)\over dt} = -k_1 N(t) \exp(-{N_0 - N(t)\over N_{kT}}),
\end{equation}
where $N_{kT} = kTC / e^2$ is the number of electrons necessary to
change the potential in the pocket by $kT/e$ and 
$k_1 = D_0 / (L\Delta L)$ is the diffusion rate.

We estimate that the pockets and barriers are a modest fraction of a
pixel size, so that $L \approx \Delta L \approx 2\mu$m, and the 
diffusivity in silicon is $D_0 \approx 30$~cm$^2$/sec, so 
$k_1^{-1} \approx 1$~nsec.  The capacitance between
the charge and the gate is $4\times10^{-8}$~F/cm$^2$, or 
$C \approx 10^{-15}$~F for these pockets.  At $T = -100$~C
we have $kT \approx 15$~meV and 
$N_{kT} \approx 100$~$e^-$.

At $t=0$ we imagine that the pocket is full of charge, $N(0) = N_0$,
and the charge then drains out.  The solution to this equation is an
exponential integral, but its approximate behavior is straightforward
to describe.
The charge in a shallow pocket with $e\Delta\phi_0 \approx kT$ will
lose its charge very quickly, but in a deep pocket with
$e\Delta\phi_0 \gg kT$ the
exponential term will dominate, and the charge will drain
more and more slowly.  For $t \ll k_1^{-1}\exp(N_0/N_{kT})$ (i.e.
$N(t) \gg N_{kT}$) we can
approximate the factor of $N$ outside the exponential as the constant $N_0$
and integrate the equation to get
\begin{equation}
N(t) \approx N_0 - N_{kT}\; \ln(1+{N_0\over N_{kT}}\,k_1 t).
\end{equation}
However, our situation calls for quantities of charge much less
than $N_{kT}$, and this last part of the draining is even simpler to
describe.  If we define $x = N(t)/N_{kT}$ as the number of electrons
in the pocket in units of $N_{kT}$, a characteristic time scale 
$\tau = \exp(N_0/N_{kT})\, k_1^{-1}$, and a dimensionless time
variable $s = t / \tau$, equation 3 becomes simply $dx / ds = -x e^x.$
In the limit that $x \ll 1$ (i.e. the last fraction of $N_{kT}$),
$e^x\approx 1$, and the solution is trivially $x(s) = e^{-s}$.  Thus
over a time scale of $\tau$ we will see a
very rapid drop of charge down to a level of $N_{kT}$, and
subsequently an exponential decay of the rest of the charge with 
time scale $\tau$: 
\begin{equation}
N(t) \approx N_{kT} \, \exp(-t/\tau) \qquad \hbox{for} \qquad t > \tau.
\end{equation}

We see a fairly uniform reduction of a factor of 30 in pocket size as
we slow the pixel dwell time from $\approx100\mu$sec to 
$\approx30$msec, which can be plugged into equation (5) to give
$1/30 \approx \exp(-30\hbox{msec}/\tau)$, or 
$k_1^{-1} \exp(N_0/N_{kT}) \approx 10$msec, or 
$N_0/N_{kT} \approx 16$.  Although the pockets have a large range of
electron capacities, this constancy indicates that the
barrier height $\Delta\phi_0$ is relatively constant,
since $N_0 / N_{kT} = e\Delta\phi_0 / kT$,
so $\Delta\phi_0 \approx 0.25$V and 
$N_0 \approx 1000 e^-$ for a typical pocket.

\subsection{Reducing the Incidence of Pockets}

We have a number of clues about the origin of the pockets we see in
the OT region of the CCD.  We find that (a) pockets occur in perhaps 1
pixel out of 300, (b) this incidence is substantially higher under
the OT gates than
under the 3-phase gates, (c) pockets are almost all unidirectional,
and horizontal trapping and vertical trapping are related in the
sense that we would expect for transfer across the $45^\circ$ boundary
between phases 1 and 2 (pockets which trap charge moving down
generally also trap charge moving left and likewise for up-right,
which is the same sense as the diagonal interface between phases 1 and 2),
(d) the pockets are very sensitive to clock
voltages, and (e) the pockets reduce by about a factor of 30 when the
dwell time in phases 1 and 2 is increased to 30~msec.  

Our best model for understanding this is that occasionally a pixel
will have an electrostatic pocket in the acute corners of phases 1 or
2 and the potential there looks like a hanging valley (i.e. steep barriers
into adjacent phases and shallow barrier into the center of the
triangle).  Charge is very
reluctant to be clocked out of this pocket into one of the
adjacent phases, which gives rise to the pocket pumping.
However, given enough time, the charge will slowly drain into the
center of the phase from which it will readily move into a different
phase.  A simple physical analysis of the situation suggests that
these pockets have a fairly constant barrier of 0.25~V, but vary
greatly in physical size, capacitance, and capacity for electrons.

The low incidence of pockets indicates that this is not a true design
flaw, but that occasionally something goes wrong and a pocket gets
created.  We are suspicious of the aluminum used for gate \#1,
since it is suspected of having small voids associated with hillock
formation after the customary forming-gas ($H_2$/$N_2$) anneal.  Processing
can also cause aluminum to lift or be undercut, but examination of
pocket-bearing pixels has not revealed this to be the case.  Any
nonuniformity in the separation between channel and gate can cause a
pocket.  Fabricating a device where the fourth level phase is
polysilicon will permit us to test this hypothesis about the origin of
the pockets.

\section{The Atmosphere and Image Motion}

\subsection{Observations}

Our first run with the OTCCD took place in July 1996, and the clouds
relented long enough to collect data on four nights.  During that time
the seeing (guided but without OT tracking) ranged from
0.63\arcsec\ to 1.27\arcsec, which is typical for the MDM 2.4-m telescope on 
Kitt Peak.

The device which we used has a rather low quantum efficiency because
it is front-side illuminated and has more than half of each pixel
covered by opaque aluminum.  We found that we would collect 1 electron
per second from a source of $m\approx 23.5$, but this will improve by
about 1.5~mag when we have a back-side illuminated, antireflection
coated OTCCD.

While removal of image motion by OT tracking is not true adaptive
optics, and although we will never rival the performance of HST, we
can obtain very significant improvements in scientific quality.
Figure 9 
\insfig{
\begin{figure}
\epsscale{1.0}
\plotone{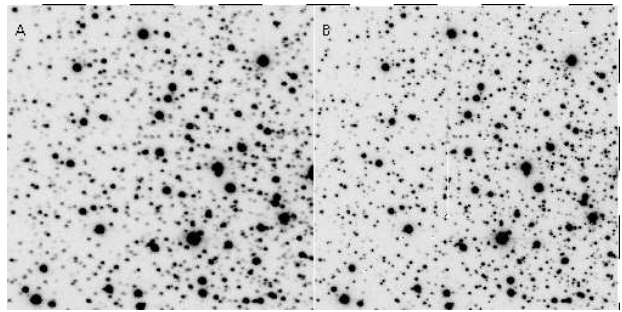}
\caption[fig9.eps]{
Two 150~sec images of M71 taken in immediate succession.  The left
one (A) used only normal guiding and has a \PSF\ \FWHM\ of 0.73\arcsec;
the right image (B) had 1113 OT shifts applied and resulting images of
0.50\arcsec.
\label{fig9}}
\end{figure}
}
illustrates the sort of improvement we achieved with OT
tracking of image motion.  The left panel shows an image of the
globular cluster M71 with ordinary guiding when the seeing was quite
good (0.73\arcsec).  The right panel shows the improvement when we
carry out OT tracking at 7~Hz (0.50\arcsec).  Although the seeing
was not always this good, we almost always achieved this level of
improvement (describable as removal of 0.5\arcsec\ in quadrature, or
reduction in \FWHM\ by 20\%) over normal guiding by use of
OT tracking.  We took a number of pairs of exposures with normal
guiding followed by OT tracking, and Figure 10 
\insfig{
\begin{figure}
\epsscale{1.0}
\plotone{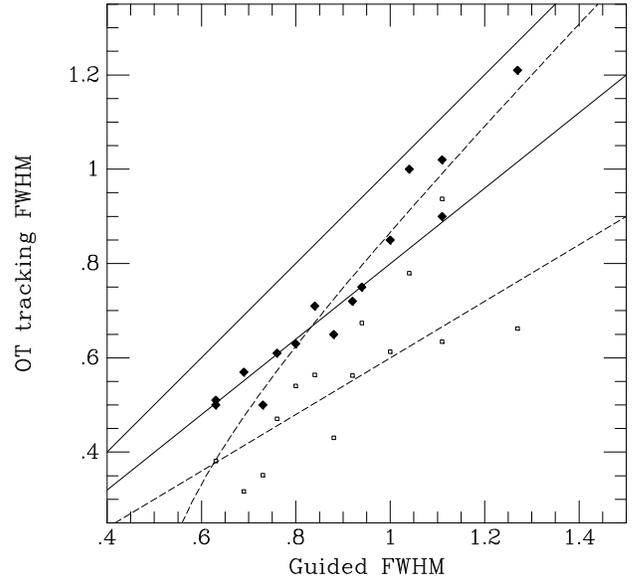}
\caption[fig10.eps]{
The \PSF\ \FWHM\ resulting from OT tracking is plotted against the
\FWHM\ from normal guiding (solid diamonds).  The \FWHM\ of the image
motion (called $\alpha_G$ below) is shown as small open squares.  The
upper solid line is $y=x$, 
the lower solid line is $y = 0.8 x$, the upper dashed line is 
$y = (x^2 - 0.5^2)^{1/2}$, and the lower dashed line is $y = 0.6 x$.
\label{fig10}}
\end{figure}
}
plots the \FWHM\
with OT tracking against the \FWHM\ with normal guiding, illustrating
this level of improvement.  Also shown on the plot is the \FWHM\ 
motion of the image centroid observed during the observation.  
The dashed line demonstrates that this motion is roughly 0.6 times
the \FWHM.

Perhaps even more remarkable, we {\it never} saw any degradation of
the \PSF\ across the OTCCD, which implies that the motion of the guide
star and science field must be very highly correlated across at least
2.5\arcmin.  Figure 11 
\insfig{
\begin{figure}
\epsscale{1.0}
\plotone{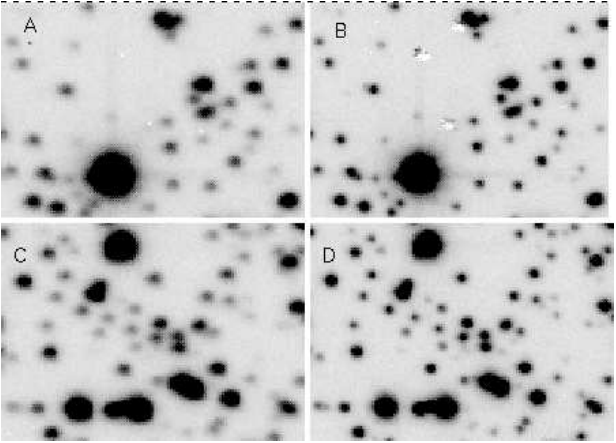}
\caption[fig11.eps]{
Two small portions of Figure 9 are shown in greater detail.  The left
images (A and C) are with ordinary tracking, the right (B and D) 
are with OT shifting, the
upper images (A and B) are about 2\arcmin\ from the guide star, and
the lower images (C and D) are about 0.5\arcmin\ from the guide star.
\label{fig11}}
\end{figure}
}
is a blowup of two regions of Figure 9, one
from the lower part of the OTCCD which was about 0.5\arcmin\ from the
guide star, and the other from the upper part at 2.0\arcmin\
separation.  Figure 12 
\insfig{
\begin{figure}
\epsscale{1.0}
\plotone{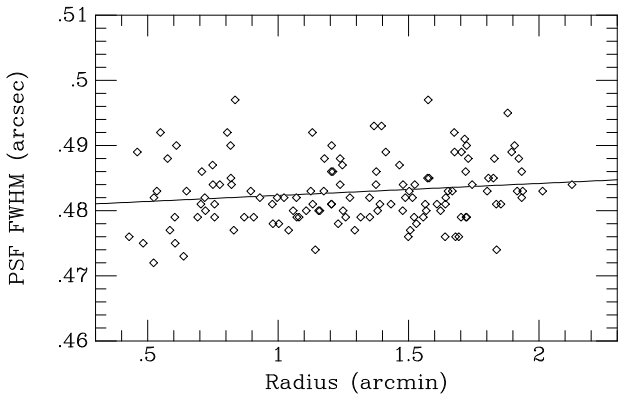}
\caption[fig12.eps]{
The \PSF\ \FWHM\ is shown as a function of distance from the guide
star.  The line is a formal fit to the data, but the data are
consistent with no variation at all.
\label{fig12}}
\end{figure}
}
quantifies this by showing that the image
\FWHM\ does not vary significantly with distance from the guide star.
The fit rises from 0.481\arcsec\ at $r=0.3$\arcmin\ to 0.485\arcsec\
at 2.3\arcmin, indicating (possibly) a contribution of 0.06\arcsec\
in image size due to uncorrelated image motion.

A further clue to the origin of the image motion comes from its
temporal behavior.  Figure 13 
\insfig{
\begin{figure}
\epsscale{1.0}
\plotone{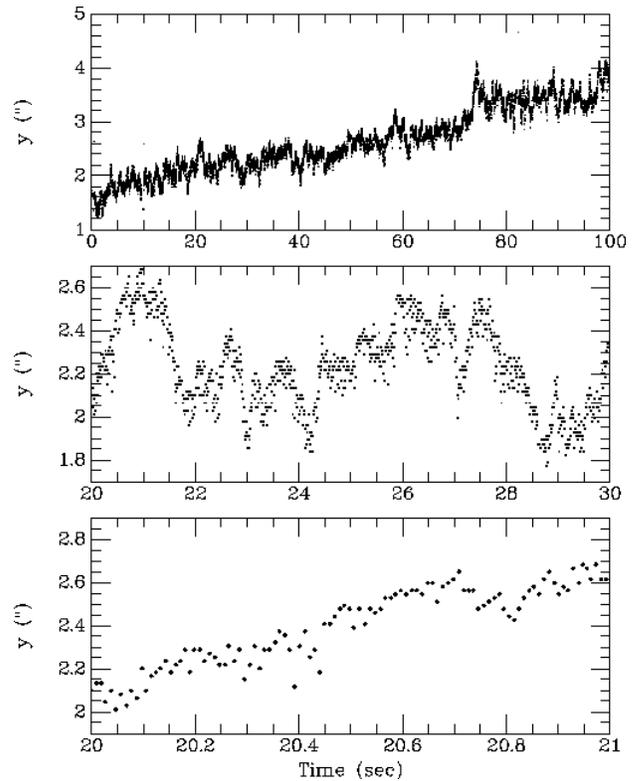}
\caption[fig13.eps]{
A time series of image motion in one direction is plotted against time
for three different time intervals.  The greatest power occurs on
time scales of roughly 1~sec.
\label{fig13}}
\end{figure}
}
shows a time series of (unguided) image
motion in one direction.  The top and bottom panels demonstrate that
there is little power on time scales of 10~sec and 0.1~sec, whereas
the middle panel shows a great deal of power on time scales of 1~sec.
This is summarized in Figure 14 
\insfig{
\begin{figure}
\epsscale{1.0}
\plotone{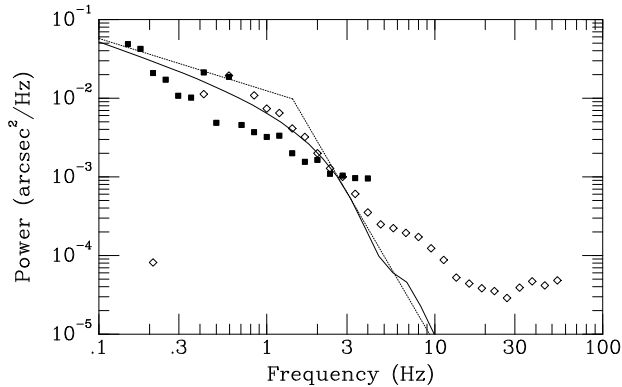}
\caption[fig14.eps]{
The power spectra of 1-axis image motion (with contributions from
positive and negative frequencies summed) is plotted against frequency
for two observations.  The solid points were derived in the $I$ band
during seeing of about 0.7\arcsec\ (note the spikes at 0.4 and 0.6
Hz caused by autoguider error), and the open points were $V$ band
observations in 0.85\arcsec\ seeing.  The centroiding has an \RMS\
error of about 0.1 pixel which causes a noise floor around
$10^{-4}$~arcsec$^2$/Hz.  The curves are from an actual integration of
the theoretical power spectrum using a Hufnagel-Valley model for
$C_n^2$ scaled to 
0.7\arcsec\ seeing along with a Bufton wind profile (solid line), and
the broken power law approximation to it (dotted line).  Most of the
power occurs at $f\approx1$~Hz where the power spectrum becomes
steeper than $f^{-1}$.  (The abcissa is log~Hz, but the ordinate is
power per Hz, not per log~Hz.)
\label{fig14}}
\end{figure}
}
as a power spectrum.

\subsection{Kolmogorov Turbulence}

The structure function of the atmospheric turbulence which causes
image distortion must be integrated over the entire atmosphere, and
the net amount of distortion 
is usually characterized in terms of the Fried parameter
$r_0 = [0.423 (2\pi/\lambda)^2 \mu_0]^{-3/5}$, where $\mu_0$ 
is the 0$^{th}$ moment of $C_n^2$:
$\mu_m = \sec^{m+1}\xi \int dh\, h^m\, C_n^2(h)$ ($\xi$ is the
zenith angle).  The \FWHM\ of the
resulting image is then approximately $\alpha = \lambda / r_0$
(e.g. Racine 1996).
Two other quantities of interest are the 1-axis tilt variance, given
by $\sigma_G^2 = 0.5 \times 5.675\, D^{-1/3}\, \mu_0$, and its power 
spectrum $S(f)$ (Sasiela and Shelton 1993, and Tyler 1994).
Combining the expressions for $\alpha$ and $\sigma_G^2$, evaluating at
$\lambda = 650$~nm, and expressing the results in terms of arcseconds,
we find that the 1-axis \FWHM\ of the image motion, $\alpha_G$, can be
expressed in terms of the overall \FWHM\ as
$$\alpha_G = 0.61\,
\left(\lambda \over 650\hbox{nm}\right)^{1/6}\,
\left(2.4\hbox{m}\over D\right)^{1/6}\,
\alpha^{5/6}.$$
This is shown on Figure 10, and is not a bad match to the observed
image motion.  Subtracting this in quadrature is likewise not a bad
match to the OT-tracking \FWHM, although this may be somewhat
fortuitous because we believe there is a contribution of about
0.4\arcsec\ to the \FWHM\ from astigmatism.

The dependence of image motion on telescope aperture is remarkably
weak.  A 10-m telescope will have image motion whose \FWHM\ is 0.5
of the overall seeing.  The frequency will be lower than a 2.4-m
telescope by a factor of 4, which makes it less obvious to the eye and
more possible for an autoguided telescope to follow it.
On the other hand it is possible that a 10-m telescope could 
do worse because of its greater moment of inertia.

Sasiela and Shelton also calculate the tip/tilt variance between two
stars as a function of separation angle, and this is illustrated in
Figure 15 
\insfig{
\begin{figure}
\epsscale{1.0}
\plotone{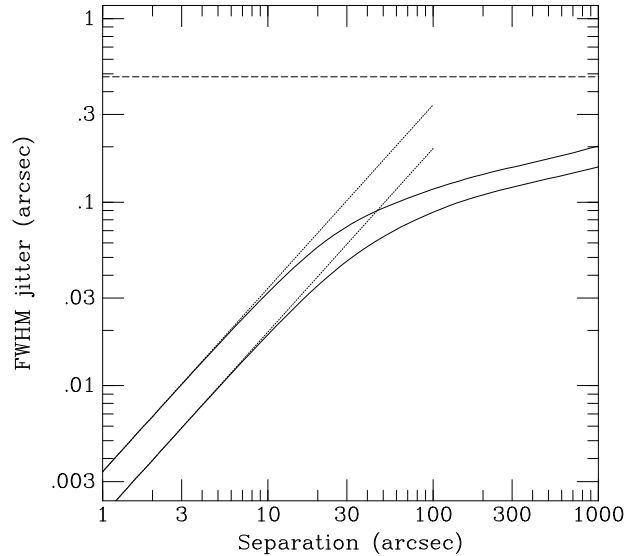}
\caption[fig15.eps]{
The theoretical \FWHM\ 1-axis image motion (\FWHM\ $\approx 2.4$ \RMS)
of one star which is being shifted in accordance with the motion of
another star, as a function of the separation of the stars.  This uses
a HV model for $C_n^2$ which has been scaled to 0.5\arcsec\ seeing.
The upper curve is the motion in the radial direction between the
stars and the lower curve is the tangential motion.  The horizontal, dashed
asymptote is $2^{1/2}$ times the jitter of one star.  Note that
the actual curves deviate very strongly on arcminute scales 
from the usual power law approximations which are accurate for small
angles of separation (dotted lines).
\label{fig15}}
\end{figure}
}
for a Hufnagel-Valley (HV; see Sasiela and Shelton Appendix
B for references and formulae) parametrization of $C_n^2$ which has
been scaled to give \FWHM$ = 0.5\arcsec$.  The eventual asymptote
is $2.35 * 2^{1/2} * 0.145\arcsec$, square root of 2 times 
the \FWHM\ motion of a single star, $\alpha_G$, 
and correcting one star's motion by the other will
improve the image as long as the differential motion is less than half
of this.  This calculation predicts differential motion at a
separation of 2\arcmin\ of 0.12\arcsec\ \FWHM\ in the perpendicular
direction and 0.15\arcsec\ in the parallel direction, which would lead
to images of $0.495\times0.505\arcsec$ at the edge of the chip.  This is
a more rapid degradation than is seen in Figure 12, and suggests that
the turbulence occurs lower than the standard HV profile.

Finally, we can calculate the expected power spectrum for our
observations, which depends not only on $C_n^2$ but also on the wind
profile $v(h)$, under the ``frozen turbulence'' assumption, whereby
the wind blows a stationary pattern of phase distortions across the
telescope aperture.  It is not difficult to match the power spectrum
reasonably well by adjusting the wind profile (we use the standard
Bufton (1973) profile, with a ground wind speed of 10~m/s), as seen in Figure
14.  For this wind profile, half of the variance occurs at
frequencies less than $f = 0.2$~Hz.  Of course, in order to reduce
the tilt jitter to less than the diffraction limit of the telescope,
it is necessary to operate at much higher frequencies, but this demonstrates
that even fairly slow tracking can remove most of the atmospheric
image motion.  Unfortunately most autoguiders have effective 3dB
points at a small fraction of a Hz, partly because of slow sample rates
but mainly because of the large inertia of the telescope.
A typical time scale for a telescope to get up to speed
after an abrupt application of the drive motors is about 1~sec.  
This implies that $\omega_{3dB} \approx 2$~sec$^{-1}$, and 
$f_{3dB} \approx 0.3$~Hz.  Without careful attention to the control
loop, such a system will do a rather poor job of removing the image
motion, even as slow as it is.

\section{Conclusions}

We have designed and built a new type of CCD which is capable of
tracking image motion by shifting the developing electronic image to
follow it.  These first devices are $512\times512$ with an equal
sized guide star area, which translates to a 1.5\arcmin\ field at the
telescope.  Characterization of these devices in the lab and
observations at the MDM 2.4-m telescope indicate that they are indeed
very effective at removing image motion.

The amplifiers achieve 1.6$e^-$ noise at a responsivity of
16$\mu$V/$e^-$.  The charge transfer efficiency is good enough (better
than 0.999993) that a 
normal exposure consisting of thousands of shifts will not
significantly degrade the \PSF.  The biggest concern about the
effectiveness of these devices was the expectation that traps and
pockets would dig holes in an image.  However, this has proven not be
be a serious problem.  We do find that pockets are about three times
more numerous under the OT gates than the three-phase gates, and we
are suspicious that the use of aluminum for one of the gates has
caused this enhancement in pockets.

Our observations using the device occurred during four nights when the
seeing ranged from 0.6\arcsec\ to 1.3\arcsec.  We find that use of OT tracking
sharpens images to approximately 80\% of their normal size, or
alternatively, OT tracking removes about 0.5\arcsec\ in quadrature
from the image \FWHM.  The \FWHM\ of image motion is roughly 60\% of
the \FWHM\ of the \PSF, and this is in good agreement with the
Kolmogorov theory of turbulence.  This theory predicts that image
motion should be a very weak function of telescope aperture and its 
frequency goes inversely as telescope aperture (under the assumption
of fixed wind speed from site to site), so removal of
image motion is important for all telescopes.

We could match the power spectra of image motion quite well with
Kolmogorov theory based on models of $C_n^2$ and the wind profile, and
we found that the image motion is surprisingly slow.  Most of the power
occurs at frequencies of less than 1~Hz, but autoguiders moving
a massive telescope may not have the bandwidth to remove much of this
motion.  However, OT tracking at 10~Hz or less removes virtually all
of the image motion.

In contrast to Kolmogorov theory and these models, however, we
observed no degradation of image quality as a function of distance
from the guide star.  The present device only permits us a range of
about 2.5\arcmin, but the models which successfully matched the power
spectrum predict a fall-off in image quality which we did not
observe.  Therefore we believe that much of the contribution to our
seeing comes from lower elevations, perhaps a boundary layer or even
within the dome, and we expect that removal of image motion should be
effective to quite large angles from the guide star, perhaps as large
as 10\arcmin.

Our development of a $512\times512$ OTCCD was based on the necessity
to fit within the chords of wafers destined for another project, and
on the expectation that image motion would become uncorrelated beyond
about 2\arcmin.  Our hope for the future is to build a larger device,
perhaps $2048\times2048$ with adjoining guide star regions of
$512\times2048$ because of the convenience and simplicity of being
able to perform the entire job of sensing and compensating for star 
motion with one standard set of CCD electronics.

\acknowledgments

This research was supported by NSF grant AST-9314665.

\endinsfig

\clearpage

\figcaption[fig1.eps]{
On the left is a schematic of how the four gates of an OTCCD are
organized.  On the right is a photomicrograph of an OTCCD pixel.  In
this device gate \#1 is aluminum and therefore appears bright.
\label{fig1}}

\figcaption[fig2.eps]{
The OTCCD is used to track image motion by placing the science field
on the portion of the CCD with OT gates and a guide star on the
frame-store portion of the CCD.  A subarray containing the star image
is rapidly read out and used to correct the position of the charge
within the OT array to follow the motion of the science field.
\label{fig2}}

\figcaption[fig3.eps]{The four pixels of a quadrant sensor.
\label{fig3}}

\figcaption[fig4.eps]{
The read noise (declining curve) and responsivity (rising curve) of
the OTCCD output amplifier as a function of drain voltage.
\label{fig4}}

\figcaption[fig5.eps]{
The number of pockets as a function of the pocket depth (determined by
pocket pumping at a rate of 80$\mu$sec per shift).  The squares
and diamonds refer to pockets which trap charge when it is clocked
up-down or left-right.
\label{fig5}}

\figcaption[fig6.eps]{
Two images of NGC~6703.  The left image (A) was taken with a Loral
CCD using normal guiding, and had an equivalent exposure of 5000~sec.
The right image (B) had 4650 shifts during a 500~sec
exposure.  The \PSF\ \FWHM\ is 0.82\arcsec\ on the left and
0.58\arcsec\ on the right.  The stretch is adjusted to make the
pockets stand out.
\label{fig6}}

\figcaption[fig7.eps]{
The size of pockets in images derived during actual observing is
plotted against the size of the same pockets derived from pocket
pumping experiments in the lab.  The line illustrates a factor of 30
reduction in size.
\label{fig7}}

\figcaption[fig8.eps]{
A pocket of volume $L * A$ has a potential barrier of
height $\Delta\phi_0$ when the pocket is empty and width
$\Delta L$.  Electons must overcome the barrier $\Delta \phi$ in order
to escape from the pocket.
\label{fig8}}

\figcaption[fig9.eps]{
Two 150~sec images of M71 taken in immediate succession.  The left
one (A) used only normal guiding and has a \PSF\ \FWHM\ of 0.73\arcsec;
the right image (B) had 1113 OT shifts applied and resulting images of
0.50\arcsec.
\label{fig9}}

\figcaption[fig10.eps]{
The \PSF\ \FWHM\ resulting from OT tracking is plotted against the
\FWHM\ from normal guiding (solid diamonds).  The \FWHM\ of the image
motion (called $\alpha_G$ below) is shown as small open squares.  The
upper solid line is $y=x$, 
the lower solid line is $y = 0.8 x$, the upper dashed line is 
$y = (x^2 - 0.5^2)^{1/2}$, and the lower dashed line is $y = 0.6 x$.
\label{fig10}}

\figcaption[fig11.eps]{
Two small portions of Figure 9 are shown in greater detail.  The left
images (A and C) are with ordinary tracking, the right (B and D) 
are with OT shifting, the
upper images (A and B) are about 2\arcmin\ from the guide star, and
the lower images (C and D) are about 0.5\arcmin\ from the guide star.
\label{fig11}}

\figcaption[fig12.eps]{
The \PSF\ \FWHM\ is shown as a function of distance from the guide
star.  The line is a formal fit to the data, but the data are
consistent with no variation at all.
\label{fig12}}

\figcaption[fig13.eps]{
A time series of image motion in one direction is plotted against time
for three different time intervals.  The greatest power occurs on
time scales of roughly 1~sec.
\label{fig13}}

\figcaption[fig14.eps]{
The power spectra of 1-axis image motion (with contributions from
positive and negative frequencies summed) is plotted against frequency
for two observations.  The solid points were derived in the $I$ band
during seeing of about 0.7\arcsec\ (note the spikes at 0.4 and 0.6
Hz caused by autoguider error), and the open points were $V$ band
observations in 0.85\arcsec\ seeing.  The centroiding has an \RMS\
error of about 0.1 pixel which causes a noise floor around
$10^{-4}$~arcsec$^2$/Hz.  The curves are from an actual integration of
the theoretical power spectrum using a Hufnagel-Valley model for
$C_n^2$ scaled to 
0.7\arcsec\ seeing along with a Bufton wind profile (solid line), and
the broken power law approximation to it (dotted line).  Most of the
power occurs at $f\approx1$~Hz where the power spectrum becomes
steeper than $f^{-1}$.  (The abcissa is log~Hz, but the ordinate is
power per Hz, not per log~Hz.)
\label{fig14}}

\figcaption[fig15.eps]{
The theoretical \FWHM\ 1-axis image motion (\FWHM\ $\approx 2.4$ \RMS)
of one star which is being shifted in accordance with the motion of
another star, as a function of the separation of the stars.  This uses
a HV model for $C_n^2$ which has been scaled to 0.5\arcsec\ seeing.
The upper curve is the motion in the radial direction between the
stars and the lower curve is the tangential motion.  The horizontal, dashed
asymptote is $2^{1/2}$ times the jitter of one star.  Note that
the actual curves deviate very strongly on arcminute scales 
from the usual power law approximations which are accurate for small
angles of separation (dotted lines).
\label{fig15}}


\clearpage

\begin{thebibliography}{}

\bibitem[Beckers 1993]{Bec93}
 Beckers, J.M. 1993, \araa, 31, 13.

\bibitem[Burke et al. 1994]{bur94}
 Burke, B.E., Reich, R.K., Savoye, E.D., \& Tonry, J.L. 1994,
 IEEE Trans. Electron Devices, 41, 2482.

\bibitem[Bufton 1973]{buf73}
 Bufton, J.L. 1973, Appl. Opt. 12 1785.

\bibitem[Chatterjee et al.]{cha79}
Chatterjee, P.K., Taylor, G.W., \& Tasch, A.F. 1979, IEEE Trans. on
Electron Devices, ED-26, 6, 871.

\bibitem[Cuillandre et al 1992]{Cui92}
   Cuillandre, J.-C., Fort, B., Picat, J.P., Soucail, 
   G., Altieri, B., Beigbeder, F., Dupin, J.P., Pourthie T., Ratier, G., 
   1994, \aap, 281, 603.

\bibitem[Jacoby et al. 1992]{Jac92} Jacoby, G.H., Branch, D.,
        Ciardullo, R., Davies, R.L., Harris, W.E., Pierce, M.J., Pritchet,
        C.J., Tonry, J.L., Welch D.L. 1992, \pasp, 104, 599.

\bibitem[Kolmogorov. 1941]{kol41}
 Kolmogorov, A.N. 1941, Dan. S.S.S.R., 30(4), 229.

\bibitem[McCarthy et al. 1997]{mcc97} 
 McCarthy, J.K., Lu, B.W., Butcher, B.A., Behr, B.B., Jensen-Grey, S.,
 Stubbs, C., Jim, K., \& Brown, Y. 1997, \pasp, in preparation.

\bibitem[McClure et al. 1989]{HRCAM} McClure, R.,  Grundmann, W.A.,
 Rambold, W.N., Fletcher, J.M., Richardson, E.H., Stillburn, J.R., Racine, R.,
 Christian, C.A., \& Waddell, P.  1989, \pasp, 101, 1156.

\bibitem[Metzger, M.R., Tonry, J.L., and Luppino, G.A.]{met93}
 Metzger, M.R., Tonry, J.L., \& Luppino, G.A. 1993,
    ASP. Conf. Ser. 52, 300.

\bibitem[Parenti and Sasiela 1993]{par94}
  Parenti, R.R. \& Sasiela, R.J. 1994,  JOSAA, 11, 288.

\bibitem[Press et al. 1992]{NR92}
  Press, W.H., Teukolsky, S.A., Vetterling, W.T., \& Flannery, B.P.
  1992, {\it Numerical Recipes}, Second edition, 
  Cambridge University Press: Cambridge

\bibitem[Racine 1996]{Rac96}
  Racine, R. 1996, \pasp, 108, 699.

\bibitem[Rigaut et al. 1991]{COMEON}
 Rigaut, F., Kern, P., Lena, P., Rousset, G., Fontanella, J.C.,
 Gaffard, J.P., \& Merkle, F.  1991, \aap, 250, 280.

\bibitem[Sasiela and Shelton 1993]{Sas93}
Sasiela, R.J. \& Shelton, J.D. 1993, JOSAA, 10, 646.

\bibitem[Schechter et al]{Sch97}
Schechter, P.L. et al. 1997, \apjl, 475, L85.

\bibitem[Sembach \& Tonry 1996]{Sem96}
 Sembach, K.R. \& Tonry, J.L. 1996, \aj, 112, 797.

\bibitem[Stockman 1982]{Sto82}
 Stockman, H.S. 1982, SPIE Proc 331, 76.

\bibitem[Tatarski 1961]{Tat61}
 Tatarski, V.I. 1961, ``Wave Propagation in a Turbulent Medium'',
 Dover:New York.

\bibitem[Thompson and Ryerson 1984]{ISIS}
 Thompson, L. \& Ryerson, H.R. 1984, Proc. SPIE, 445, 560.

\bibitem[Tonry et al.]{ton97}
 Tonry, J.L., Blakeslee, J.P., Ajhar, E.A., Dressler A., 1997, \apj, 475, 399.

\bibitem[Tyler 1994]{Tyl94}
 Tyler, G.A. 1994, JOSAA, 11,358.

\end{thebibliography}
\end{document}